\documentclass[onecolumn,preprintnumbers,eqsecnum,superscriptaddress,10pt,floatfix,amsmath,amssymb]{revtex4}
\usepackage{float}
\usepackage{graphicx}
\usepackage{epsf}
\usepackage{dcolumn}   % Align table columns on decimal point
\usepackage{bm}        % bold math
\usepackage{color}
\usepackage[titletoc]{appendix}
\usepackage[colorlinks,citecolor=blue]{hyperref}
\usepackage[marginal]{footmisc}
\newcommand{\be}{\begin{equation}}
\newcommand{\en}{\end{equation}}
 \newcommand{\bea}{\begin{eqnarray}}
 \newcommand{\ena}{\end{eqnarray}}
  
\begin{document}

\title{Investigating strong gravitational lensing with black hole metrics modified with an additional term}

\author{Xiao-Jun Gao}\email{xjgao2020@nuaa.edu.cn}
\address{College of Science, Nanjing University of Aeronautics and Astronautics, Nanjing 210016, China}

\author{Ji-Ming Chen} \email{chenjm94@mail.ustc.edu.cn}
\address{College of Science, Nanjing University of Aeronautics and Astronautics, Nanjing 210016, China}

\author{Hongsheng Zhang}\email{sps$_$zhanghs@ujn.edu.cn}
\address{School of Physics and Technology, University of Jinan, 336 West Road of Nan Xinzhuang, Jinan, Shandong 250022, China}

\author{Yihao Yin}\email{yinyihao@nuaa.edu.cn}
\address{College of Science, Nanjing University of Aeronautics and Astronautics, Nanjing 210016, China}
\address{Key Laboratory of Aerospace Information Materials and Physics (NUAA), MIIT, Nanjing 211106, China}

\author{Ya-Peng Hu}\email{huyp@nuaa.edu.cn}
\address{College of Science, Nanjing University of Aeronautics and Astronautics, Nanjing 210016, China}
\address{Key Laboratory of Aerospace Information Materials and Physics (NUAA), MIIT, Nanjing 211106, China}

\begin{abstract}
Gravitational lensing is one of the most impressive celestial phenomena, which has interesting behaviors in its strong field limit. Near such limit, Bozza finds that the deflection angle of light is well-approximated by a logarithmic term and a constant term. In this way he explicitly derived the analytic expressions of deflection angles for a few types of black holes. In this paper, we study the explicit calculation to two new types of metrics in the strong field limit: (i) the Schwarzschild metric extended with an additional $r^{-n}(n\geq 3)$ term in the metric function; (ii) the Reissner-Nordstrom metric extended with an additional $r^{-6}$ term in the metric function. With such types of metrics, Bozza's original way of choosing integration variables may lead to technical difficulties in explicitly expressing the deflection angles, and we use a slightly modified version of Bozza's method to circumvent the problem.

%PACS number: 04.70.Dy, 04.20.-q, 04.50.Kd

\end{abstract}

\maketitle

%\vspace*{1.cm}

%\newpage

\section{Introduction}
As an important gravitational effect of general relativity, Einstein in 1916 predicted that the light from a star passing around a compact object will be deflected~\cite{Einstein:1916}. This prediction was confirmed by astronomical observation in 1919, and since then has passed tests with higher and higher accuracy \cite{Eddington:1919,Lebach:1995zz,Shapiro:2004zz}. This phenomenon is similar to that of light passing through optical lenses, and hence is called gravitational lensing (GL),
%In fact, different from optical lenses, the effect of gravitational lensing can produce multiple images from a single light source. For example, the Einstein ring is formed by the light source MG1131+0456~\cite{Hewitt:1988} due to GL, this case only is the light source, compact object and observer all in the same direction.
and it has become an important tool in astrophysics. One can distinguish naked singularities from black holes by investigating positions, magnifications and time delays of relativistic images resulting from GL\cite{Virbhadra:2002ju,Virbhadra:2007kw,Gyulchev:2008ff,Sahu:2012er}. Furthermore, GL can also be used to test different theories of gravity\cite{Keeton:2005jd,Keeton:2006sa}.

Through gravitational lensing, the light source Q0957+561 forms two quasar images \cite{Walsh:1979nx}, while MG1131+0456 forms Einstein ring \cite{Hewitt:1988}. These categories of gravitational lensing are often called the strong gravitational lensing, which usually involves multiple images and high magnifications of the light source. In the strong GL, there are weak and strong gravitational field regions, where the trajectory of light can present different behaviors. These regions are also known to correspond to the weak and strong field limit during calculations like the deflection angle of light, and here the weak and strong limit are usually dependent on the distance between the light and photon sphere. For example, the well-known deflection angle predicted by Einstein in 1916 and many early works were done in the weak field limit \cite{Sereno:2003nd,Keeton:2005jd,Gao:2019pir}. On the other hand, the strong field limit has also been theoretically investigated since decades ago with many interesting results \cite{Virbhadra:1999nm,Frittelli:1999yf,Eiroa:2002mk,Bozza:2002af,Vazquez:2003zm,Bozza:2003cp}. In Schwarzschild black hole lensing, Virbhadra and Ellis found a sequence of relativistic images on both sides of the optic axis apart from the primary and the secondary images~\cite{Virbhadra:1999nm}, while Fritelli, Kling and Newman obtained the exact expressions for magnifications and time delays of relativistic images~\cite{Frittelli:1999yf}. Torres obtained analytic expressions for positions and magnifications of relativistic images in Reissner-Nordstr\"om black hole lensing~\cite{Eiroa:2002mk}. Bozza investigated quasiequatorial GL by spinning black holes, and derived formulas of positions and magnifications of their relativistic images~\cite{Bozza:2002af}. V$\acute{a}$zquez and Esteban researched observational properties of the relativistic image for the Kerr black hole lensing with an arbitrary source and observer~\cite{Vazquez:2003zm}. Bozza and Mancini demonstrated that the time delay between the first two images in black hole lensing is proportional to the minimum impact angle, and this ratio gives a precise measure of the distance between the observer and the black hole~\cite{Bozza:2003cp}.

Bozza has also found that, with a general metric that is static, spherically symmetric and asymptotically flat, the deflection angle exhibits a logarithmic divergence in the strong field limit~\cite{Bozza:2001xd,Bozza:2002zj}. Near such limit, the analytic expression of the deflection angle is well-approximated by the sum of a logarithmic term and a constant term. Besides the general discussion, Bozza further specifically investigated a few examples (the Schwarzschild, Reissner-Nordstr\"om, and Janis-Newman-Winicour black holes) with explicit results. Many further results on specific metrics have also been obtained by using the same method as Bozza's. For some recent examples, see ~\cite{Tsukamoto:2016jzh,Shaikh:2019jfr,Lu:2019ush,Zhu:2020wtp,Tsukamoto:2020uay,Konoplya:2020hyk}. To calculate the deflection angle, integrals over the radial coordinate $r$ have to be done, and a technical detail should be noted that, in Bozza's method, there is a change in the integration variable from $r$ to $z$ (details explained in Sec.\ref{Bozzamethod}), where $z(r)$ is always explicitly defined. However, depending on the particular metric in question, the explicit expression of the inverse function $r(z)$ can be difficult to be obtained or may even do not exist at all, which in practice hinders the derivation of the deflection angle.

In this paper, we investigate the following two types of spherically symmetric metrics with an additional term in the metric function: (i) the Schwarzschild metric extended with $r^{-n}\ (n\geq 3, n \in \mathbb{R})$ ; (ii) the Reissner-Nordstr\"om metric extended with $r^{-6}$. We are interested in these metrics because: (a) as far as we know, in the literature they have not yet been discussed in the context of strong field limit, but they may be of interest to future research in different types of black holes, and in particular, the one with $r^{-6}$ is physically motivated by ~\cite{Maceda:2018zim,Magos:2020ykt}; (b) the extra $r^{-n}$ terms make it difficult or even impossible to have explicit $r(z)$ as mentioned above, which urges us to improve detailed techniques when applying Bozza's method.

Our paper is organized as follows. In Sec.\ II, we first do a brief review of Bozza's work on the logarithmic behavior of the light deflection angle in the strong field limit. In Sec.\ III, following the spirit of Bozza's work, by slightly modifying his technique of doing integrals, we present the mathematical framework of how to obtain the analytic expression of the deflection angle. Then in Sec.\ IV we explicitly investigate the deflection angles in the Schwarzschild metric extended with an additional $r^{-3}$ or $r^{-5}$ term and in the Reissner-Nordstr\"om metric extended with an additional $r^{-6}$ term. In the last section, we make conclusions and briefly discuss possible future research directions. Throughout this paper we use the geometric units with $G=c=1$.

\section{Review: Logarithmic behavior of the deflection angle in the strong field limit}
\label{Bozzamethod}
In this section, we briefly make a review of Ref.\cite{Bozza:2002zj} on the logarithmic behavior of the light deflection angle in the strong field limit. For a four-dimensional spherically symmetric static spacetime, its general line element is given by
\begin{align}
 {\rm d}s^2=-A(r){\rm d}t^2+B(r){\rm d}r^2+C(r)({\rm d}\theta^2+\sin^2\theta{\rm d}\varphi^2). \label{spacetime}
\end{align}
For an asymptotic flat spacetime, functions $A(r), B(r)$ and $C(r)$ have the following asymptotic behaviors
\begin{align}\label{eq:2.2}
\lim \limits_{r \to \infty}A(r)=1, \quad \lim \limits_{r \to \infty}B(r)=1,  \quad \lim \limits_{r \to \infty}C(r)=r^2.
\end{align}
The equation of the photon sphere is given by \cite{Claudel:2000yi}
\begin{align}
\dfrac{C'(r)}{C(r)}=\dfrac{A'(r)}{A(r)}, \label{PhotonSphere}
\end{align}
where the prime denotes differentiation with respect to the radial coordinate $r$. The largest root of Eq. (\ref{PhotonSphere}) $r_m$ is called the radius of the photon sphere \cite{Claudel:2000yi}, while $A(r), B(r)$ and $C(r)$ are often positive for $r\geq r_m$.

If this spacetime (\ref{spacetime}) is generated by a compact object, i.e. like a black hole, the light will be deflected when passing around.
Taking the spherical symmetry into account, we only investigate, without loss of generality, the light traveling on the equatorial plane of this compact object, i.e. $\theta=\pi/2$. The trajectory of the light, as a null geodesic, is given as \cite{Hu:2014}
\begin{align}
\dfrac{{\rm d}r}{{\rm d}\varphi}=\sqrt{\frac{C(r)}{B(r)}}\sqrt{\dfrac{C(r)}{A(r)}\dfrac{1}{u^2}-1}, \label{trajectory}
\end{align}
where $u\equiv \dfrac{L}{E}$ is a constant called the impact parameter, and $E$ and $L$ are the conserved energy and angular momentum, respectively,  of the light. One can use ${\rm d}r/{\rm d}\varphi=0$ to obtain the minimum distance $r_0$ between the light's trajectory and the compact object, which satisfies
\begin{align}
u=\sqrt{\dfrac{C(r_0)}{A(r_0)}}.\label{r0}
\end{align}
This equation gives the relation between the two important parameters $u$ and $r_0$. From trajectory (\ref{trajectory}), one obviously finds that the light has been deflected, and the corresponding deflection angle is expressed as
\begin{align}\label{deflection angle}
\hat{\alpha}(r_0)=I(r_0)-\pi,
\end{align}
where
\begin{align}
I(r_0)\equiv\int^{\infty}_{r_0} \dfrac{2\sqrt{B(r)}}{\sqrt{C(r)} \sqrt{\dfrac{C(r)}{A(r)}\dfrac{A(r_0)}{C(r_0)}-1}}{\rm d}r.\label{IntegralFunction}
\end{align}
This deflection angle has been found to diverge logarithmically when $r_0$ approaches the photon sphere radius $r_m$~\cite{Bozza:2002zj}. In the following, we will briefly review how to explicitly express this logarithmic behavior. For convenience of discussion, Bozza in Ref.\cite{Bozza:2002zj} did a change of variable:
\begin{align}
z= \dfrac{A(r)-A(r_0)}{1-A(r_0)},\label{transformation}
\end{align}
and the coordinate $r$ is evaluated by
\begin{equation}
	r=A^{-1}[(1-A(r_0))z+A(r_0)]. \label{invfunc}
\end{equation}
The function $I(r_0)$ in Eq.(\ref{IntegralFunction}) is hence rewritten as
\begin{align}\label{eq:2.9}
I(r_0)=\int^{1}_{0}R(z,r_0)f(z,r_0){\rm d}z,
\end{align}
where
\begin{align}\label{eq:2.10}
R(z,r_0)\equiv\dfrac{2\sqrt{B(r)A(r)}}{C(r)A'(r)}[1-A(r_0)]\sqrt{C(r_0)},\quad f(z,r_0)\equiv\dfrac{1}{\sqrt{A(r_0)-[(1-A(r_0))z+A(r_0)]\dfrac{C(r_0)}{C(r)}}},
\end{align}
Note that, $R(z,r_0)$ is regular, while $f(z,r_0)$ diverges when $z\rightarrow 0$, i.e. equivalent to $r\rightarrow r_0$. The divergent part of $f(z,r_0)$, namely $f_D(z,r_0)$, is expressed as $f_D(z,r_0)=1/\sqrt{\alpha(r_0) z+\beta(r_0) z^2}$, where $\alpha(r_0)$ and $\beta(r_0)$ are given \cite{Bozza:2002zj}
\begin{align}
\alpha(r_0)=&\frac{1-A(r_0)}{C(r_0)A'(r_0)}[C'(r_0)A(r_0)-C(r_0)A'(r_0)],\label{bozzaalpha}\\
\beta(r_0)=&\frac{[1-A(r_0)]^2}{2C(r_0)^2A'(r_0)^3}\big[2C(r_0)C'(r_0)A'(r_0)^2+[C(r_0)C''(r_0)-2C'(r_0)^2]A(r_0)A'(r_0)-C(r_0)C'(r_0)A(r_0)A''(r_0)\big].\label{bozzabeta}
\end{align}

Then the function $I(r_0)$ can be divided into a divergent part $I_D(r_0)$ and a regular part $I_R(r_0)$
\begin{align}
I_D(r_0)=&\int^{1}_{0}R(0,r_0)f_D(z,r_0){\rm d}z,\label{DivergentPart}\\
I_R(r_0)=&\int^{1}_{0}[R(z,r_0)f(z,r_0)-R(0,r_0)f_D(z,r_0)]{\rm d}z.\label{RegularPart}
\end{align}
The divergent function $I_D(r_0)$ can be further explicitly integrated as
\begin{align}\label{12}
I_D(r_0)=\dfrac{2R(0,r_0)}{\sqrt{\beta(r_0)}}\log{\dfrac{\sqrt{\beta(r_0)}+\sqrt{\alpha(r_0)+\beta(r_0)}}{\sqrt{\alpha(r_0)}}}.
\end{align}
From this explicit result, one easily sees the logarithmic behavior of $I_D(r_0)$ as $r_0$ approaches $r_m$
\begin{align}\label{eq:2.12}
I_D(r_0)=-a\log(\dfrac{r_0}{r_m}-1)+b_D+O(r_0-r_m),
\end{align}
where
\begin{align}\label{eq:2.13}
a=&\dfrac{R(0,r_m)}{\sqrt{\beta(r_m)}}, \quad b_D=\dfrac{R(0,r_m)}{\sqrt{\beta(r_m)}}\log{\dfrac{2[1-A(r_m)]}{r_mA'(r_m)}}, \\
\beta(r_m)=&\dfrac{C(r_m)[1-A(r_m)]^2[C''(r_m)A(r_m)-C(r_m)A''(r_m)]}{2A^2(r_m)C'^2(r_m)}.\label{Bozzabetarm}
\end{align}
The regular function $I_R(r_0)$ gives a constant term $b_R$ on the photon sphere, i.e.\ $b_R=I_R(r_m)$. Thus, in the strong field limit, i.e.\ when $r_0$ is near $r_m$ and gravity is strong, the deflection angle $\hat{\alpha}(r_0)$ is expanded as
\begin{align}
\hat{\alpha}(r_0)=-a\log{(\dfrac{r_0}{r_m}-1)}+b_D+b_R-\pi+O(r_0-r_m),\label{Deflectionangle}
\end{align}
where the first term on the right hand side shows the logarithmic behavior. Generally, the angular position $\vartheta$ of the light source image, which can be directly measured in astronomical observations, is related to the impact parameter $u$ as $u=\vartheta D_{OL}$, and $D_{OL} $ is the distance between the compact object and the observer. Therefore, it is more convenient to rewrite the deflection angle (\ref{Deflectionangle}) as a function of $u$ by using (\ref{r0}) \cite{Bozza:2002zj}
\begin{align}
\hat{\alpha}(u)=-\bar{a}\log{(\dfrac{u}{u_m}-1)}+\bar{b}+O(u-u_m),\label{DeAngle-u}
\end{align}
where
\begin{align}
u_m=\sqrt{\dfrac{C(r_m)}{A(r_m)}},~~\bar{a}=\dfrac{R(0,r_m)}{2\sqrt{\beta(r_m)}},~~\bar{b}=\bar{a}\log{\dfrac{2\beta(r_m)}{A(r_m)}}+b_R-\pi.\label{urm}
\end{align}

\section{Deflection angle in the strong field limit in a generic framework}
\label{modmethod}
In this section, we give the deflection angle in the strong field limit in a generic framework, using a slightly modified version of Bozza's method \cite{Bozza:2002zj}. Note that, with the change of variable (\ref{transformation}), whether we can explicitly express the deflection angle (\ref{DeAngle-u}) depends on whether we can write the right hand side of (\ref{invfunc}) as an explicit function, i.e.\ express $r$ explicitly in terms of $z$. Unfortunately, the explicit expressions are difficult to obtain, if the $A(r)$ in the metric function contains terms like $r^{-n}$ for $n\geq3$, or may be even impossible for $n\geq5$. To avoid this difficulty, we do the integrals in a way slightly different from Bozza's.

We start from the integral Eq.(\ref{IntegralFunction}), rewriting it as
\begin{align}
I(r_0)=\int^{\infty}_{r_0}\bar{R}(r,r_0)\bar{f}(r,r_0){\rm d}r, \label{IntegralFunctionNew}
\end{align}
where
\begin{align}
\bar{R}(r,r_0)\equiv\dfrac{2\sqrt{B(r)A(r)}}{C(r)A'(r)}\sqrt{C(r_0)}, \quad \bar{f}(r,r_0)\equiv\dfrac{A'(r)}{\sqrt{A(r_0)-A(r)\dfrac{C(r_0)}{C(r)}}}.\label{newRterm}
\end{align}
Obviously, the function $\bar{R}(r,r_0)$ is also regular, while $\bar{f}(r,r_0)$ is divergent when $r\rightarrow r_0$. Following a procedure similar to the previous section, the divergent part of $\bar{f}(r,r_0)$, namely $\bar{f}_D(r,r_0)$, is expressed as
\begin{align}\label{eq:3.3}
\bar{f}_D(r,r_0)=\dfrac{A'(r)}{\sqrt{\bar{\alpha}(r_0)[A(r)-A(r_0)]+\bar{\beta}(r_0)[A(r)-A(r_0)]^2}},
\end{align}
where
\begin{align}
\bar{\alpha}(r_0)=&\dfrac{1}{C(r_0)A'(r_0)}\left[A(r_0)C'(r_0)-A'(r_0)C(r_0)\right], \label{alphar0}\\
\bar{\beta}(r_0)=&\dfrac{1}{2C(r_0)A'^2(r_0)}\left[A(r_0)C''(r_0)-A''(r_0)C(r_0)\right].\label{betar0}
\end{align}
By using $\bar{f}_D(r,r_0)$, the $I(r_0)$ in (\ref{IntegralFunctionNew}) is divided into divergent part $\bar{I}_D(r_0)$ and regular part $\bar{I}_R(r_0)$, which are
\begin{align}
\bar{I}_D(r_0)=&\int^{\infty}_{r_0}\bar{R}(r_0,r_0)\bar{f}_D(r,r_0){\rm d}r,\label{eq:3.6}\\
\bar{I}_R(r_0)=&\int^{\infty}_{r_0}[\bar{R}(r,r_0)\bar{f}(r,r_0)-\bar{R}(r_0,r_0)\bar{f}_D(r,r_0)]{\rm d}r, \label{RegularR0}
\end{align}
where $R(r_0,r_0)=R(r,r_0)\mid_{r=r_0}$. For the divergent part $\bar{I}_D(r_0)$, it can be further explicitly integrated as
\begin{align}\label{DivergentID}
\bar{I}_D(r_0)=\dfrac{2\bar{R}(r_0,r_0)}{\sqrt{\bar{\beta}(r_0)}}\log{\dfrac{\sqrt{\bar{\beta}(r_0)}\sqrt{1-A(r_0)}+\sqrt{\bar{\alpha}(r_0)+\bar{\beta}(r_0)[1-A(r_0)]}}{\sqrt{\bar{\alpha}(r_0)}}}.
\end{align}
Note that, $\bar{\alpha}(r_0)$ vanishes at $r_0=r_m$ since $\bar{\alpha}(r_0)$ in (\ref{alphar0}) is related to equation of photon sphere (\ref{PhotonSphere}). Therefore, $\bar{I}_D(r_0)$ is obviously divergent at $r_0=r_m$. After expanding $\bar{\alpha}(r_0)$ around $r_0=r_m$ as
\begin{align}
\bar{\alpha}(r_0)=2A'(r_m)\bar{\beta}(r_m)(r_0-r_m)+O[(r_0-r_m)^2],\label{expandingalphar0}
\end{align}
where $\bar{\beta}(r_m)$ is just the above coefficient of $\bar{f}_D(r,r_0)$ in (\ref{betar0})
\begin{align}
\bar{\beta}(r_m)=\bar{\beta}(r_0)\mid_{r_0=r_m}=\dfrac{1}{2A'^2(r_m)C(r_m)}[C''(r_m)A(r_m)-A''(r_m)C(r_m)],\label{newbeita}
\end{align}
and then substituting (\ref{expandingalphar0}) into (\ref{DivergentID}), we obtain
\begin{align}\label{eq:3.12}
\bar{I}_D(r_0)=-a'\log(\dfrac{r_0}{r_m}-1)+b'_D+O(r_0-r_m),
\end{align}
where
\begin{align}\label{eq:3.13}
a'=\dfrac{\bar{R}(r_m,r_m)}{\sqrt{\bar{\beta}(r_m)}},~~b'_D=\dfrac{\bar{R}(r_m,r_m)}{\sqrt{\bar{\beta}(r_m)}}\log{\dfrac{2[1-A(r_m)]}{r_mA'(r_m)}},
\end{align}
and $\bar{R}(r_m,r_m)=\bar{R}(r,r_0)\mid_{r,r_0=r_m}$. For the regular part $\bar{I}_R(r_0)$, the constant term is given by $b'_R=\bar{I}_R(r_m)$. To obtain the analytic expression of $b'_R$, it is more convenient to do a change of variable $x=r_0/r$ in (\ref{RegularR0}), which gives
\begin{align}
\bar{I}_R(r_0)=\int^{1}_{0}[\bar{R}(x,r_0)\bar{f}(x,r_0)-\bar{R}(r_0,r_0)\bar{f}_D(x,r_0)]\dfrac{r_0}{x^2}{\rm d}x,\label{Trick}
\end{align}
and this trick will be useful to actually calculate this term in the following applications for some black hole spacetimes. Thus, the deflection angle $\hat{\alpha}(r_0)$ in (\ref{deflection angle}) is finally expressed as
\begin{align}
\hat{\alpha}(r_0)=-a'\log(\dfrac{r_0}{r_m}-1)+b'+O(r_0-r_m), \label{FinalAlphar0}
\end{align}
where $b'=b'_R+b'_D-\pi$. If we use the impact parameter $u$ as the variable instead of $r_0$, the above formula is further rewritten as
\begin{align}
\hat{\alpha}(u)=-\bar{a}'\log(\dfrac{u}{u_m}-1)+\bar{b}'+O(u-u_m), \label{FinalAlphau}
\end{align}
where
\begin{align}\label{eq:3.18}
\bar{a}'=\dfrac{\bar{R}(r_m,r_m)}{2\sqrt{\bar{\beta}(r_m)}}, \quad \bar{b}'=\bar{a}'\log{\dfrac{2\bar{\beta}(r_m)[1-A(r_m)]^2}{A(r_m)}}+b'_R-\pi.
\end{align}
and
\begin{align}
b'_R=\bar{I}_R(r_m)\equiv \int^{1}_{0} \chi(x,r_m){\rm d}x=\int^{1}_{0}[\bar{R}(x,r_m)\bar{f}(x,r_m)-\bar{R}(r_m,r_m)\bar{f}_D(x,r_m)]\dfrac{r_m}{x^2}{\rm d}x,\label{Trick1}
\end{align}
with $\chi(x,r_m)\equiv [\bar{R}(x,r_m)\bar{f}(x,r_m)-\bar{R}(r_m,r_m)\bar{f}_D(x,r_m)]\dfrac{r_m}{x^2}$. In Appendix~\ref{A}, as a cross-check we demonstrate that our result of the deflection angle (\ref{FinalAlphau}) is consistent to Bozza's result (\ref{DeAngle-u}) in the Reissner-Nordstr\"om black hole spacetime.

\section{Black hole metrics with an additional $r^{-n}\ (n\geq 3)$ term}
For the two simplest types of spherically symmetric black holes, i.e. the Schwarzschild and Reissner-Nordstr\"om black holes, Bozza in \cite{Bozza:2002zj} has given the explicit expressions of the deflection angles in the strong field limit. For the purpose of investigating larger varieties of black holes in various theories of gravity, we are motivated to extend the investigation to a bit more generalized situations. In this section, using the method introduced above, we will calculate a few examples where an additional $r^{-n}\ (n\geq 3)$ term is added to the metric function of Schwarzschild or Reissner-Nordstr\"om. In the following, for simplicity we will work in the units that the black hole mass parameter $M =\frac{1}{2}$ (or equivalently that the Schwarzschild radius is set to 1).

\subsection{The Schwarzschild metric extended with an additional $r^{-3}$ or $r^{-5}$ term}
In this subsection we work out two examples that one extra term is added to the Schwarzschild metric function. As a warm-up exercise, let us first try the simplest example with an additional $r^{-3}$ term
\begin{align}
 A(r)=B(r)^{-1}=1-\dfrac{1}{r}-\dfrac{F}{r^3}, \quad C(r)=r^2,\label{Fspacetime}
\end{align}
where $F$ is a constant parameter, and the event horizon locates at $r=r_+$ which is the largest root of $A(r)=0$. In this example, an additional $r^{-3}$ term is introduced, and consequently the explicit expression of $r(z)$, although obtainable from (\ref{transformation}), is technically too complicated for doing the integral (\ref{RegularPart}). Thus here we follow the calculation in Sec.\ \ref{modmethod} instead without involving the variable $z$.
We easily obtain that $F$ should satisfy $F\geq-4/27$ to keep the existence of horizon. For the $F\geq 0$ case, we find that $A(r)=0$ just has one positive root, i.e. one horizon, while there are two horizons for the $-4/27<F<0$ case. In Fig.1 and Fig.2, we have plotted function $A(r)$ in these two cases with $F=0.776$ and $F=-0.111$, respectively.
\begin{figure}[htbp]
\begin{minipage}[t]{0.48\linewidth}
\centering
\includegraphics[width=3.2in]{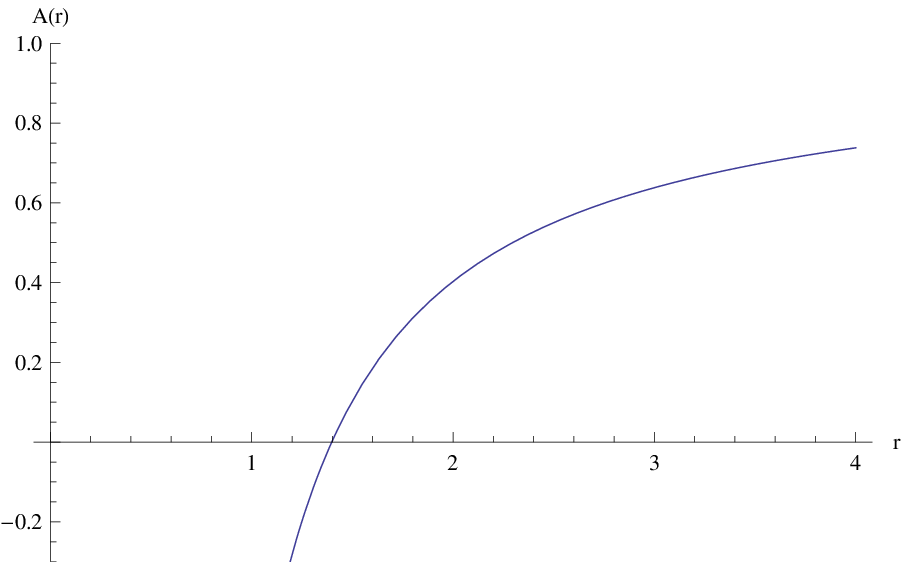}
\caption{The function $A(r)$ plotted in the case $F=0.776$.}
\label{fig:side:a}
\end{minipage}
\hspace{0.1cm}
\begin{minipage}[t]{0.48\linewidth}
\centering
\includegraphics[width=3.2in]{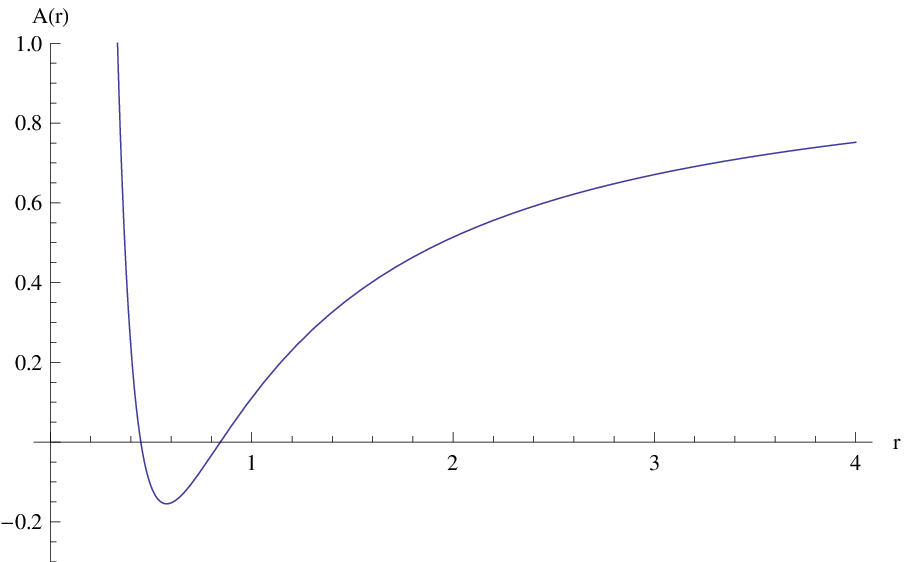}\\
\caption{The function $A(r)$ plotted in the case $F=-0.111$.}
\label{fig:side:b}
\end{minipage}
\end{figure}

The radius of photon sphere is easily obtain from (\ref{PhotonSphere})
\begin{align}
r_m=\frac{1}{2} \left(1+\frac{1}{\sqrt[3]{10 F+2 \sqrt{5} \sqrt{F (5 F+1)}+1}}+\sqrt[3]{10 F+2 \sqrt{5} \sqrt{F (5 F+1)}+1}\right),\label{PhotonSphereF}
\end{align}
while the corresponding impact parameter $u_m$ is
\begin{align}
u_m=\sqrt{\dfrac{C(r_m)}{A(r_m)}}=\sqrt{\frac{{r_m}^5}{r_m^2 (r_m-1)-F}}.\label{Fimpactparameter}
\end{align}

In order to obtain the strong field limit coefficients of the deflection angle in (\ref{FinalAlphau}) for this static spacetime, we first calculate the corresponding three constants $\bar{R}(r_m,r_m)$, $\bar{\beta}(r_m)$ and $b'_R$. By substituting (\ref{Fspacetime}) into (\ref{newRterm}), (\ref{newbeita}) and (\ref{Trick}), we easily obtain
\begin{align}
&\bar{R}(r_m,r_m)=\dfrac{2\sqrt{B(r_m)A(r_m)}}{C(r_m)A'(r_m)}\sqrt{C(r_m)}=\frac{2 r_m^3}{3 F+ r_m^2},\label{RrmA3}\\
&\bar{\beta}(r_m)=\dfrac{1}{2A'^2(r_m)C(r_m)}[C''(r_m)A(r_m)-A''(r_m)C(r_m)]=\frac{r_m^3 (5 F+r_m^3)}{(3 F+ r_m^2)^2},\label{betaA3}
\end{align}
and
\begin{align}
b'_R=&\bar{I}_R(r_m)=\int^{1}_{0}[\bar{R}(x,r_m)\bar{f}(x,r_m)-\bar{R}(r_m,r_m)\bar{f}_D(x,r_m)]\dfrac{r_m}{x^2}{\rm d}x\notag\\
=&2 \log \left[6 (2-\sqrt{3})\right]+\frac{8[-4 \sqrt{3}+18-5 \log 6+10 \coth ^{-1}\left(\sqrt{3}\right)]}{27}F+O(F^2),\label{IrmA3}
\end{align}
where we do the analytic calculation only to the first order of $F$. Therefore, we obtain the strong field limit coefficients of the deflection angle (\ref{FinalAlphau}) for this static black hole spacetime as
\begin{align}
\bar{a}'=\frac{r_m^{3/2}}{\sqrt{5 F+r_m^3}},~~\bar{b}'=\bar{a}'\log \left[-\frac{2 (5 F+r_m^3)(F+ r_m^2)^2}{(3 F+ r_m^2)^2 (F+ r_m^2-r_m^3)}\right]+b'_R-\pi.\label{Finallylimitcofficient}
\end{align}

In order to easily read the first-order correction from the $F$, we further write the final expression of the deflection angle as the consistent power-series in terms of the coefficient $F$. After substituting (\ref{PhotonSphereF}), (\ref{Fimpactparameter}) and (\ref{Finallylimitcofficient}) into (\ref{FinalAlphau}), then expanding (\ref{FinalAlphau}) in powers of $F$, we finally obtain the expression of the deflection angle in the strong field limit as the consistent power-series in terms of the coefficient $F$
\begin{align}
\hat{\alpha}_{F}(u)=&-\left(1-\frac{20 F}{27}\right)\log\left[u-\dfrac{3\sqrt{3}}{2}-\dfrac{2\sqrt{3}F}{3}\right]+\log[324(7 \sqrt{3}-12)]-\pi\notag\\
&+\frac{2}{27} \left[66-16 \sqrt{3}-30 \log6-5 \log \left(\frac{27}{4}\right)+40 \coth ^{-1}\sqrt{3}\right]F+O(F^2).\label{Fdeflectionangle}
\end{align}
Obviously, the above result with $F=0$ returns to that of the Schwarzschild black hole case in \cite{Bozza:2002zj}. A subtlety needs to be clarified here that the first divergent logarithm term is unsuitable to further expand in terms of $F$. The reason is that the further expansion is unphysical since the radius of photon sphere should contain the modification from the small coefficient $F$. Obviously, when $u=\frac{3\sqrt{3}}{2}$, which is the value corresponding to the photon sphere of the Schwarzschild black hole case, this logarithmic term does not diverge due to the corrections from $F$. On the other hand, if we expand the logarithm in powers of $F$, every term in the power series explodes when $u$ approaches $\frac{3\sqrt{3}}{2}$.

It should be also emphasized that, in the above $n=3$ example, the calculation relies on the expression (\ref{PhotonSphereF}) of $r_m(F)$ as an exact solution to (\ref{PhotonSphere}). Such an exact solution also exists for $n=4$, but can be impossible to obtain for $n\geq5$.
Furthermore we would like to mention again that, the application of Bozza's original method, which involves explicit $r(z)$ of (\ref{invfunc}), is feasible up to $n=4$, so our modified method does not show much advantage, unless we go for $n\geq5$.
Therefore, considering that the $n=4$ example shows nothing new, but merely another page of complicated formulas, we would like to skip it and jump to $n=5$ directly.

Now we investigate the $n=5$ example as
\begin{align}
A(r)=B(r)^{-1}=1-\dfrac{1}{r}-\dfrac{P}{r^5}, \quad C(r)=r^2,\label{Sspacetime}
\end{align}
where $P$ is a constant parameter, which needs to satisfy $P\geq-256/3125$ in order to make sure that the horizon exists. $A(r)=0$ has only one positive root for $P\geq 0$, and has two positive roots for $-256/3125<P<0$, corresponding to the existence of one and two horizons, respectively. In Fig.3 and Fig.4, we plot $A(r)$ with $P=0.428$ and $P=-0.041$ for illustration.
\begin{figure}[H]
\begin{minipage}[t]{0.48\linewidth}
\centering
\includegraphics[width=3.2in]{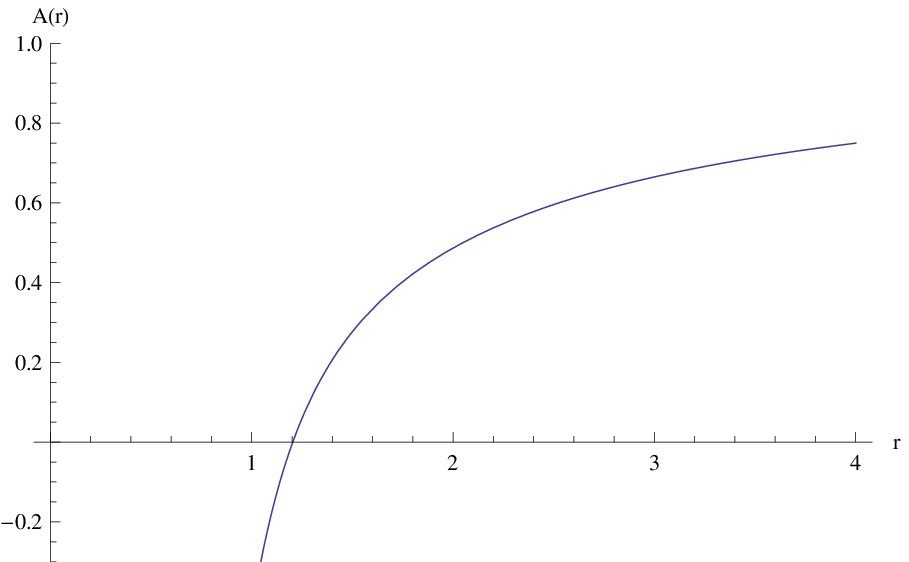}
\caption{The function $A(r)$ plotted in the case $P=0.428$.}
\label{fig:side:a}
\end{minipage}
\hspace{0.1cm}
\begin{minipage}[t]{0.48\linewidth}
\centering
\includegraphics[width=3.2in]{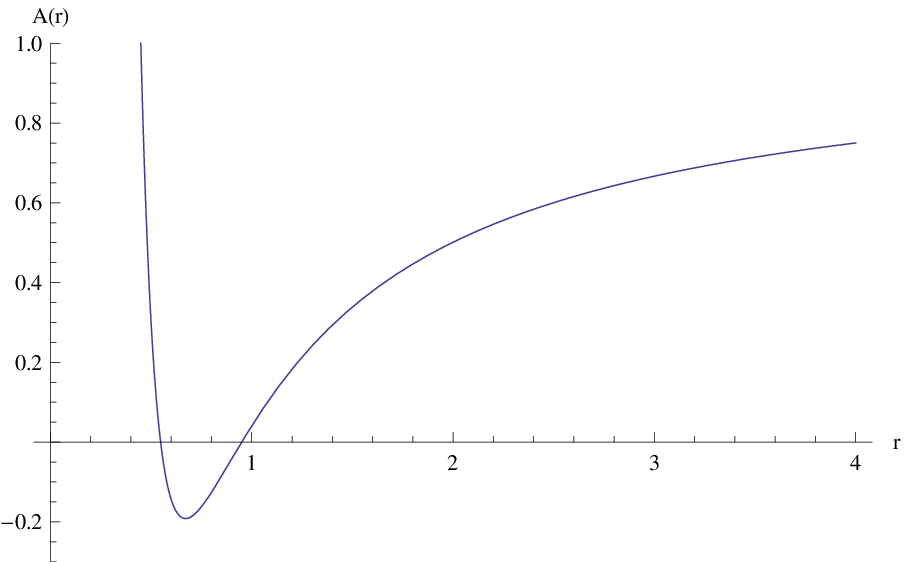}\\
\caption{The function $A(r)$ plotted in the case $P=-0.041$.}
\label{fig:side:b}
\end{minipage}
\end{figure}
The equation of the photon sphere (\ref{PhotonSphere}) for (\ref{Sspacetime}) is
\begin{align}
7 P-2 r_m^5+3 r_m^4=0,\label{A5photonsphere}
\end{align}
which is essentially a fifth degree polynomial equation. Note that, unlike (\ref{PhotonSphereF}) in the previous example, this equation may lack the algebraic solution $r_m(P)$, but this does not prevent us from proceeding with our calculation, as shown in the following.

From (\ref{A5photonsphere}), we have
\begin{align}
	\frac{\rm d}{{\rm d} P}r_m(P)=\frac{7}{10 r_m^4(P)-12 r_m^3(P)}.\label{A5rmPderive}
\end{align}
Furthermore we also obtain from (\ref{A5photonsphere}) that
\begin{align}
r_m(0)=\frac{3}{2},\label{Schwarzschildphsp}
\end{align}
which, with $P$ set to zero, is the photon sphere radius of Schwarzschild black hole.
Then together with (\ref{A5rmPderive}) we get
\begin{align}
r'_m(0)=&\frac{\rm d}{{\rm d} P}r_m(P)\big|_{P=0}=\frac{56}{81}.\label{drmP}
\end{align}
Now we calculate the strong field limit coefficients. By substituting (\ref{Sspacetime}) into (\ref{newRterm}), (\ref{newbeita}) and (\ref{Trick}), we easily obtain
\begin{align}
&\bar{R}(r_m,r_m)=\dfrac{2\sqrt{B(r_m)A(r_m)}}{C(r_m)A'(r_m)}\sqrt{C(r_m)}=\frac{2 r_m^5}{5 P+ r_m^4},\label{RrmA5}\\
&\bar{\beta}(r_m)=\dfrac{1}{2A'^2(r_m)C(r_m)}[C''(r_m)A(r_m)-A''(r_m)C(r_m)]=\frac{r_m^5 (14 P+r_m^5)}{(5 P+ r_m^4)^2},\label{betaA5}
\end{align}
and
\begin{align}
b'_R=&\bar{I}_R(r_m)=\int^{1}_{0}[\bar{R}(x,r_m)\bar{f}(x,r_m)-\bar{R}(r_m,r_m)\bar{f}_D(x,r_m)]\dfrac{r_m}{x^2}{\rm d}x=\int^{1}_{0}\chi[x,r_m(P)]{\rm d}x\notag\\
=&\int^{1}_{0}\big[\chi[x,r_m(0)]+\chi[x,r_m(0),r'_m(0)]P+O(P^2)\big]{\rm d}x\notag\\
=&2 \log \left[6 (2-\sqrt{3})\right]+\frac{32[-382 \sqrt{3}+1759-490 \log [6 \left(2-\sqrt{3}\right)]]}{8505}P+O(P^2).\label{IrmA5}
\end{align}
Here by Taylor-expanding the integrand in $P$, we obtain (\ref{IrmA5}) without involving any algebraic expression of $r_m(P)$.
Then we obtain the strong field limit coefficients in the deflection angle (\ref{FinalAlphau}) for this static black hole spacetime as
%\begin{align}
%\bar{a}'=\frac{2r_m^{3/2}[F+(2M-r_m)r_m^2]}{(3F+2Mr_m^2)\sqrt{5 F+r_m^3}},~~\bar{b}'=\bar{a}'\log \left[-\frac{(5 F+r_m^3) (F+2 M r_m^2)^2}{2 [F+r_m^2 (2 M-r_m)]^3}\right]+b'_R-\pi.\label{Finallylimitcofficient}
%\end{align}
\begin{align}
u_m=\sqrt{\frac{r_m^7}{r_m^4 (r_m-1)-P}},~~\bar{a}'=\frac{r_m^{5/2}}{\sqrt{14 P+r_m^5}},~~\bar{b}'=\bar{a}'\log \left[\frac{2 (14 P+r_m^5)(P+ r_m^4)^2}{(5 P+ r_m^4)^2 ( r_m^5-r_m^4-P)}\right]+b'_R-\pi,\label{FinallylimitcofficientA5}
\end{align}
Similarly, we substitute (\ref{FinallylimitcofficientA5}) into (\ref{FinalAlphau}), and then by using (\ref{Schwarzschildphsp}) and (\ref{drmP}) we can expand (\ref{FinalAlphau}) in powers of $P$. Thus we can also obtain the expression of the deflection angle in the strong field limit as the consistent power-series in terms of the coefficient $P$
\begin{align}
\hat{\alpha}_{P}(u)=&-\left(1-\frac{224 P}{243}\right)\log\left[u-\dfrac{3\sqrt{3}}{2}-\frac{8\sqrt{3} P}{27 }\right]+\log [324(7 \sqrt{3}-12)]-\pi\notag\\
&+\dfrac{16}{8505}\left[3483-764 \sqrt{3}-1225 \log3-980 \log [6 (2-\sqrt{3})]\right]P+O(P^2).\label{Pdeflectionangle}
\end{align}

\subsection{The Reissner-Nordstr\"om metric extended with an additional $r^{-6}$ term}
In this subsection, we will study a bit more complicated example, which is the Reissner-Nordstr\"om metric extended with an additional $r^{-6}$ term:
\begin{align}
	A(r)=B(r)^{-1}=1-\dfrac{1}{r}+\dfrac{q}{r^2}-\dfrac{\mu q^2}{20 r^6}, \quad C(r)=r^2.\label{Euler-Heisenbergspacetime}
\end{align}
This is a spherically symmetric and static solution in the Einstein-Euler-Heisenberg theory, where $q$ is related to the electric charge parameter, and $\mu$ is the Euler-Heisenberg parameter~\cite{Maceda:2018zim,Magos:2020ykt}.
When the parameters are properly chosen, horizons may exist. If $\mu>0.1$ and $0<q\leq0.25$, $A(r)=0$ has only one positive root representing the event horizon, and if $0<\mu\leq0.1$ and $0<q\leq0.25$, $A(r)=0$ has three positive roots, whose largest one is the event horizon~\cite{Allahyari:2019jqz}. For example, we have plotted $A(r)$ in Fig.5 and Fig.6, when $\mu$ is equal to $0.1872$ and $0.0264$ with $q=0.2148$, respectively.
\begin{figure}[H]
\begin{minipage}[t]{0.48\linewidth}
\centering
\includegraphics[width=3.2in]{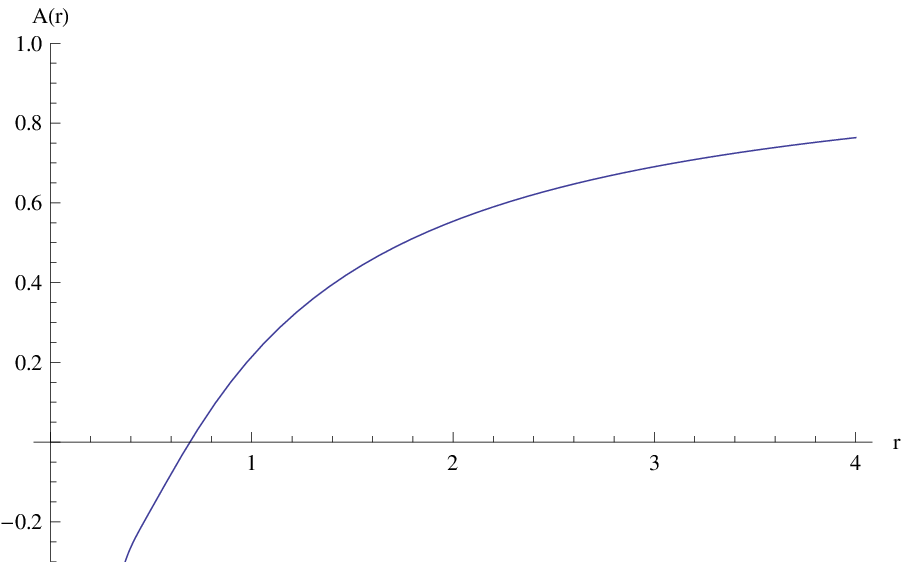}
\caption{The function $A(r)$ plotted in the case $\mu=0.1872$ and $q=0.2148$.}
\label{fig:side:a}
\end{minipage}
\hspace{0.1cm}
\begin{minipage}[t]{0.48\linewidth}
\centering
\includegraphics[width=3.2in]{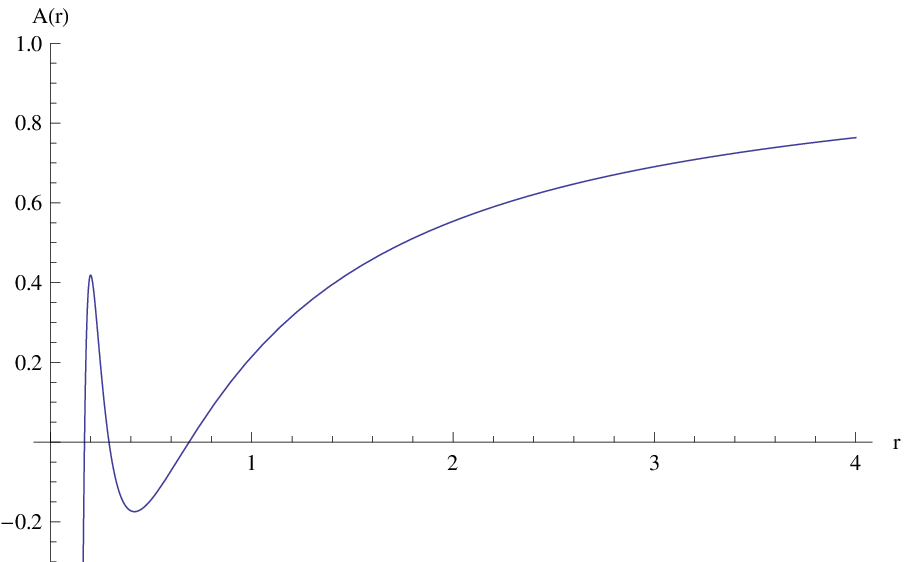}\\
\caption{The function $A(r)$ plotted in the case $\mu=0.0264$ and $q=0.2148$.}
\label{fig:side:b}
\end{minipage}
\end{figure}

By substituting (\ref{Euler-Heisenbergspacetime}) into (\ref{PhotonSphere}), the equation of the photon sphere
\begin{align}
2 \mu  q^2-20 q r_m^4-10 r_m^6+15 r_m^5=0,\label{A6potonsphere}
\end{align}
is essentially of a sixth degree polynomial, and may lack the algebraic solution $r_m(q,\mu)$. From (\ref{A6potonsphere}) we have
\begin{align}
\frac{\rm d}{{\rm d} \mu}r_m(q,\mu)=\frac{2q^2}{-75r_m^4(q,\mu)+80 q r_m^3(q,\mu)+60 r_m^5(q,\mu)}, \label{A6rmmuderive}
\end{align}
and also that
\begin{align}
r_m(q,0)=\frac{1}{4}(\sqrt{9-32 q}+3)\label{A6RNphotonsphere},
\end{align}
which is the photon sphere radius of Reissner-Nordstr\"om black hole for $\mu=0$. By substituting (\ref{A6RNphotonsphere}) into (\ref{A6rmmuderive}) we have
\begin{align}
r'_m(q,0)=\frac{\rm d}{{\rm d} \mu}r_m(q,\mu)|_{\mu=0}=\frac{512 q^2}{5 (\sqrt{9-32 q}+3)^3 (-32 q+3 \sqrt{9-32 q}+9)},\label{appendixRNrm}
\end{align}
where the prime stands for the derivative with respect to $\mu$.

The GL in the weak field limit of this type of black holes has been studied in \cite{Fu:2021akc}. In the following we will do the investigation in the strong field limit. Similar to the previous examples, by substituting (\ref{Euler-Heisenbergspacetime}) into (\ref{newRterm}), (\ref{newbeita}) and (\ref{Trick}), we first obtain the three constants $\bar{R}(r_m,r_m)$, $\bar{\beta}(r_m)$ and $b'_R$
 \begin{align}
&\bar{R}(r_m,r_m)=\dfrac{20 r_m^6}{10 r_m^4 (r_m-2q)+3 \mu  q^2},\label{barEEH}\\
&\bar{\beta}(r_m)=\dfrac{100 r_m^6 (\mu  q^2-2 q r_m^4+r_m^6)}{[10 r_m^4 (r_m-2q)+3 \mu  q^2]^2},\label{betaEEH}
\end{align}
and
\begin{align}
b'_R=&\bar{I}_R(r_m)=\int^{1}_{0}[\bar{R}(x,r_m)\bar{f}(x,r_m)-\bar{R}(r_m,r_m)\bar{f}_D(x,r_m)]\dfrac{r_m}{x^2}{\rm d}x=\int^{1}_{0}\chi[x,r_m(q,\mu)]{\rm d}x\notag\\
=&\int^{1}_{0}\big[\chi[x,r_m(q,0)]+\chi[x,r_m(q,0),r'_m(q,0)]\mu+O(\mu^2)\big]{\rm d}x\notag\\
=&2 \log \left[6 (2-\sqrt{3})\right]+\frac{8 q (\sqrt{3}-4+\log [6 (2-\sqrt{3})])}{9}+\frac{4}{81} q^2 \big[38 \sqrt{3}-169+44 \log [6(2-\sqrt{3})]\big]\notag\\
&+\frac{q^3 16\big[-1133 \sqrt{3}-7485+828 (3+\sqrt{3}) \log[6 (2-\sqrt{3})]\big]}{2187 (3+\sqrt{3})}+\frac{\mu  q^2 16\left[1521-326 \sqrt{3}-420 \log [6 (2-\sqrt{3})]\right] }{76545}\notag\\
&+\ \ (\text{terms of order} > 3).\label{IrmEEH1}
\end{align}
Here we only consider the situation that both $\mu$ and $q$ are small, and keep the terms up to the third order, where the first $\mu$-dependent term appears. As a cross-check, (\ref{IrmEEH1}) reduces to the result (\ref{IrmRN}) in the Reissner-Nordstr\"om black hole spacetime for $\mu=0$ and ignoring $q^2$ and higher order terms. Then we obtain the strong field limit coefficients of the deflection angle (\ref{FinalAlphau}) for the Einstein-Euler-Heisenberg black hole as follows
\begin{align}
u_m=&\sqrt{\frac{20 r_m^8}{20 r_m^4 (r_m^2+q- r_m)-q^2 \mu}},~~\bar{a}'=\frac{r_m^3}{\sqrt{\mu  q^2-2 q r_m^4+r_m^6}},\label{EEHimparameter}\\
\bar{b}'=&\bar{a}'\log \left[-\frac{10 (\mu  q^2-2 q r_m^4+r_m^6) [\mu  q^2-20 r_m^4 (q- r_m)]^2}{[\mu  q^2-20 r_m^4 (r_m (r_m-1)+q)] [10 r_m^4 ( r_m-2q)+3 \mu  q^2]^2}\right]+b'_R-\pi.\label{EEHbcoefficient}
\end{align}
In order to read the correction compared with that in the Schwarzschild case, here we just consider the case with both small $q$ and $\mu$. After substituting (\ref{EEHimparameter}) and (\ref{EEHbcoefficient}) into (\ref{FinalAlphau}), and then with the help of Eqs.\ (\ref{A6RNphotonsphere}) and (\ref{appendixRNrm}), we expand (\ref{FinalAlphau}) to the third power of $\mu$ and $q$, where the first $\mu$-dependent term appears. Thus we obtain the expression of the deflection angle in the strong field limit as the consistent power-series in terms of the small coefficients $q$ and $\mu$
\begin{align}
\hat{\alpha}_{EEH}(u)=&-\left(1+\dfrac{4 q}{9}+\dfrac{88 q^2}{81}+\dfrac{736 q^3}{243}-\dfrac{32 q^2\mu}{729}\right)\log\left[u-\frac{3 \sqrt{3}}{2}+\sqrt{3} q+\frac{7\sqrt{3} q^2}{9 }+\frac{302\sqrt{3} q^3}{243 }-\dfrac{4\sqrt{3} q^2\mu}{405}\right]\notag\\
&+\log[324(7 \sqrt{3}-12)]-\pi-\delta_1 q-\delta_2 q^2-\delta_3 q^3+\delta_4 q^2 \mu+\ \ (\text{terms of order} > 3),\label{mudeflectionangle}
\end{align}
where the coefficients $\delta_i$ are
\begin{align}
\delta_1=&\frac{2}{9}\left[15-4 \sqrt{3}-\log243-4 \log [6(2-\sqrt{3})]\right],\\
\delta_2=&\frac{4}{81}\left[153-38 \sqrt{3}-55 \log3-44 \log[6(2-\sqrt{3})]\right],\\
\delta_3=&\frac{16}{729(3+\sqrt{3})}\left[2205+281 \sqrt{3}-345(3+\sqrt{3}) \log3-276(3+\sqrt{3}) \log[6 (2-\sqrt{3})]\right],\\
\delta_4=&\dfrac{8}{76545}\left[3105-652 \sqrt{3}-1050 \log3-840 \log[6(2-\sqrt{3})]\right].
\end{align}

\section{Summary and discussion}
In this paper, we have first briefly made a review on the logarithmic behavior of the light deflection angle in the strong field limit, which has been investigated by Bozza in Ref.\cite{Bozza:2002zj}. Then we have calculated the explicit expressions of the deflection angles for the Schwarzschild black hole with an extra term $r^{-3}$ or $r^{-5}$ in the metric function, and for the Reissner-Nordstr\"om black hole with an extra term $r^{-6}$. Because with these two types of metrics, Bozza's original way of choosing integration variables may lead to technical difficulties in explicitly expressing the deflection angles, and we have slightly modified Bozza's method to circumvent the problem.

Note that, we have only paid attention to static black holes in this paper, and it would be interesting to further investigate strong-field deflection angles around rotating black holes. Moreover, the Event Horizon Telescope (EHT) has successfully captured the shadow image of the black hole in the center of $M87^{*}$ galaxy~\cite{Akiyama:2019cqa,Akiyama:2019eap}. Therefore, it will be interesting to investigate whether the deflection angle has also logarithmic behavior in the strong field limit of a rotating black hole, and how it imprints such shadow images. In addition, one easily finds that the strong field limit coefficients of the deflection angle is model-dependent in the context of modified gravity theories. Therefore, investigating how these coefficients affect the shadow image may give a new route to test gravitational theories. Finally, it is important to notice that, in this paper and relevant literature, the photon is usually treated as a point-like particle, but to be more precise, the nature of GL should instead be described by electromagnetic waves propagating in curved spacetime, which can lead to many interesting research in the future, e.g. how the polarization of light evolves near the strong field limit, and how non-linear electrodynamics is different from Maxwell's theory by affecting GL (the Einstein-Euler-Heisenberg theory mentioned in Sec. IV.B can be a good example to study, and the cause of the difference in GL may not be limited to the difference of metrics).

\section{Acknowledgements}
We thank a lot for the discussions with Dr. Shi-Bei Kong. This work is supported by National Natural Science Foundation of China (NSFC) under grant Nos. 12175105, 11575083, 11565017, Top-notch Academic Programs Project of Jiangsu Higher Education Institutions (TAPP), and is also supported by \textquotedblleft the Fundamental Research Funds for the Central Universities, NO. NS2020054\textquotedblright\ of China. H.S Zhang is supported by Shandong Province Natural Science Foundation under grant No. ZR201709220395, and the National Key Research and Development Program of China (No. 2020YFC2201400).

\appendix

\section{Demonstration of equivalence to Bozza's result in Reissner-Nordstr\"om spacetime} \label{A}
In this appendix, we demonstrate that our result in (\ref{FinalAlphau}) is equivalent to Bozza's result in (\ref{DeAngle-u}) in the Reissner-Nordstr\"om black hole spacetime.

The metric of Reissner-Nordstr\"om spacetime within the ansatz (\ref{spacetime}) is ~\cite{Bozza:2002zj}
\begin{align}
A(r)=1-\dfrac{1}{r}+\dfrac{q}{r^2},\quad B(r)=\left(1-\dfrac{1}{r}+\dfrac{q}{r^2}\right)^{-1}, \quad C(r)=r^2,\label{RNspacetime}
\end{align}
where we have set $M=\frac{1}{2}$, and $q$ is related to the electric charge parameter. The outer horizon of Reissner-Nordstr\"om spacetime is $r_H= \frac{1}{2} \left(\sqrt{1-4 q}+1\right)$ with $|q| \leq 1/4$. By substituting them into Eq.\eqref{PhotonSphere} and solving the equation, one easily obtains the photon sphere radius
\begin{align}
r_m=\frac{1}{4} \left(\sqrt{9-32 q}+3\right). \label{PhotonSphereRN}
\end{align}

In the following, we will apply the method in Sec.\ \ref{modmethod} to obtain the strong field limit coefficients of the deflection angle (\ref{FinalAlphau}) for Reissner-Nordstr\"om spacetime. From (\ref{FinalAlphau}), one easily obtains
\begin{align}
\bar{R}(r_m,r_m)=&\bar{R}(r,r_0)\mid_{r,r_0=r_m}=\dfrac{2\sqrt{B(r_m)A(r_m)}}{C(r_m)A'(r_m)}\sqrt{C(r_m)}=\frac{2 r_m^2}{r_m-2 q},\label{RrmRN}\\
\bar{\beta}(r_m)=&\dfrac{1}{2A'^2(r_m)C(r_m)}[C''(r_m)A(r_m)-A''(r_m)C(r_m)]=\frac{r_m^2(r_m^2-2 q)}{(r_m-2 q)^2},\label{betarmRN}
\end{align}
and
\begin{align}
b'_R=&\bar{I}_R(r_0)\mid_{r_0=r_m}=\int^{1}_{0}[\bar{R}(x,r_m)\bar{f}(x,r_m)-\bar{R}(r_m,r_m)\bar{f}_D(x,r_m)]\dfrac{r_m}{x^2}{\rm d}x\\
=&2 \log \left[6 (2-\sqrt{3})\right]+\frac{8}{9} \left[\sqrt{3}-4+\log [6 (2-\sqrt{3})]\right]q+O(q^2),\label{IrmRN}
\end{align}
where we have used the change of variable in (\ref{Trick}). Here to simplify the problem, we limit our calculation to the second order of $q$, as has also been done in Ref.~\cite{Bozza:2002zj}. Thus, we obtain the strong field limit coefficients
%\begin{align}
%\bar{a}'=\frac{2r_m(q^2+r_m^2-r_m)}{(r_m-2q^2)\sqrt{r_m^2-2 q^2}},~~\bar{b}'=\bar{a}'\log \left[\frac{(q^2-r_m)^2 (r_m^2-2 q^2)}{2 [q^2+(r_m-1) r_m]^3}\right]+b'_R-\pi,  \label{RNCoefficients}
%\end{align}
\begin{align}
u_m=\sqrt{\dfrac{r_m^4}{q+r_m(r_m-1)}},~~\bar{a}'=\frac{r_m}{\sqrt{r_m^2-2 q}},~~\bar{b}'=\bar{a}'\log \left[\frac{2 (q-r_m)^2 (r_m^2-2 q)}{(r_m-2 q)^2 (q+r_m^2- r_m)}\right]+b'_R-\pi,  \label{RNCoefficients}
\end{align}
and by substituting (\ref{PhotonSphereRN}) for $r_m$, we find that our result is consistent with Bozza's result (\ref{DeAngle-u}) in the case of Reissner-Nordstr\"om metric.
Furthermore, after substituting (\ref{RNCoefficients}) into (\ref{FinalAlphau}), we can expand (\ref{FinalAlphau}) in powers of $q$ (the first logarithmic term is not further expanded for the reason explained under (\ref{Fdeflectionangle})). Thus we obtain the expression of the deflection angle in the strong field limit as the consistent power-series in terms of the small coefficient $q$
\begin{align}
\hat{\alpha}_{RN}=&-\left(1+\frac{4 q}{9}\right)\log\left[u-\frac{3 \sqrt{3}}{2}+\sqrt{3} q\right]+\log[324(7 \sqrt{3}-12)]-\pi\notag\\
&-\frac{2}{9}\left[15-4 \sqrt{3}-\log243-4 \log [6(2-\sqrt{3})]\right]q+O(q^2).\label{RNdeflectionangle}
\end{align}

\end{document}